\documentclass[12pt]{article}
\newtheorem{theorem}{Theorem}
\newtheorem{lemma}{Lemma}
\newtheorem{proposition}{Proposition}
\newtheorem{corollary}{Corollary}

\usepackage{latexsym}

\begin{document}

\title{On the area of the symmetry orbits in $T^2$ symmetric spacetimes
with Vlasov matter}
\author{Marsha Weaver\\Theoretical Physics Institute\\
University of Alberta\\Edmonton, AB, Canada T6G 2J1\\
mweaver@phys.ualberta.ca}
\date{}
\maketitle

\begin{abstract}
This paper treats the global existence question for a collection of
general relativistic collisionless particles, all having the same mass.
The spacetimes considered are globally hyperbolic, with Cauchy surface a
three-torus.  Furthermore, the spacetimes considered are isometrically
invariant under a two-dimensional group action, the orbits of which are
spacelike two-tori.  It is known from previous work that the area of the
group orbits serves as a global time coordinate.  In the present work it
is shown that the area takes on all positive values in the maximal Cauchy
development.
\end{abstract}

\section{Introduction}
\label{intro}

Global existence results have been obtained 
\cite{Moncrief,Chr,BCIM,a1,Weav,N,ARW} for various classes of
$T^2$ symmetric spacetimes -- solutions to Einstein's equations
with a closed (compact without boundary) Cauchy surface isometrically
invariant under an effective group action, $U(1) \times U(1)$.
A key feature of these results is that the area of the
two-dimensional group orbits serves as a global time
coordinate.  In \cite{Moncrief,Chr,N} the global existence
result is obtained directly in terms of a coordinate system with areal
time coordinate.  In these cases the ``areal coordinates'' are
also ``conformal coordinates'' and the area tending to zero
signals the end of the maximal Cauchy development.\footnote{The
spatial topology is the three-torus
in \cite{Moncrief,BCIM,a1,Weav,N,ARW} and a global
time coordinate can be set equal to the area of
the group orbits.  In
\cite{Chr}, other spatial topologies are considered,
and the relationship of the area of the group orbits
to the time coordinate is specified.}
In the other cases \cite{BCIM,a1,Weav,ARW},
areal coordinates and conformal coordinates are distinct, and the
global existence argument in the contracting direction is not made
directly in terms of areal coordinates, but rather in terms of conformal
coordinates.  
(The quotient of the spacetime by
the symmetry group is conformally flat.  In conformal
coordinates the conformal factor is given explicitly.)
The results in the second group leave open the
possibility that any conformal time coordinate is unbounded in the
contracting direction and that the area tends to some positive
number (rather than all the way to zero)
at the end of the maximal Cauchy development.

In \cite{IW}, vacuum $T^2$ symmetric spacetimes such that the twist
quantities associated with the two commuting spatial Killing vector
fields do not both vanish are considered.
Areal coordinates are used directly in the contracting
direction to show that the area of the $T^2$ symmetry group orbits
tends to zero at the end of the maximal Cauchy development
(as long as the spacetime is not flat --
one can choose two Killing vector fields tangent to a Cauchy
surface in a spatially compactified
flat Kasner spacetime such that the twist quantities are not
both vanishing and such that the area of the $T^2$ group orbits
tends to a positive number at the end of the maximal Cauchy
development).  The argument used in \cite{IW} to thus sharpen
the vacuum global existence result
previously obtained in \cite{BCIM} carries over straightforwardly
to sharpen in the same way the global existence result for 
spacetimes with Gowdy symmetry ({\it i.e.,} vanishing twist
quantities) and a nonvanishing magnetic field orthogonal
to the group orbits obtained in appendix C of \cite{Weav}.
In the present work, the general strategy of \cite{IW}, using
areal coordinates directly in the contracting direction rather than
conformal coordinates, is used to so sharpen global existence
results previously obtained in \cite{a1,ARW} for
Einstein-Vlasov initial data on $T^3$ with $T^2$ symmetry.

The Einstein-Vlasov system models a general relativistic collection
of collisionless particles.  For an
introduction to the Einstein-Vlasov system, see \cite{ehlers,r1,a2,r2}.
In the present work, as in \cite{a1,ARW},
massive particles are considered, all having the same mass.
In section~\ref{preliminaries}, the areal coordinate metric
is written out, a global existence result
from \cite{ARW} is recalled, and the Vlasov matter distribution
function is introduced.  Some preliminary results pertaining
to the Vlasov matter, which are key to the presence of Vlasov
matter insuring that the area of the $T^2$ symmetry orbits
goes to zero, are obtained.  In section~\ref{ee}, Einstein's equations
are written out.  In section~\ref{extend}, bounds on the areal metric
functions are obtained in the contracting direction which are sufficient
to allow application to the contracting direction of
the techniques used in \cite{BCIM,a1,ARW} for extending the
spacetime in the expanding direction.  These techniques involve
integrating quantities related to the metric functions along the light
cones of the quotient spacetime and integrating quantities related to the
Vlasov matter along timelike geodesics.  The main theorem appears in
section~\ref{conclusion}.

There does exist an exceptional possibility for $T^2$ symmetric
spacetimes, characterized by the spacetime gradient of the area
of the symmetry group orbits vanishing on the closed Cauchy
surface, which implies that the area of the
orbits is everywhere constant, and so cannot serve as
a time coordinate.  In this exceptional case, the spacetime is flat,
either $T^3 \times R$ Minkowski spacetime or flat Kasner.  The presence
of a nonvanishing twist quantity rules out this exceptional possibility
\cite{Chr}, as does the presence of Vlasov matter (see the appendix).

\section{Preliminaries}
\label{preliminaries}

One of the results in \cite{ARW} is
that (except for the special vacuum solutions with symmetry
group orbits of constant area mentioned in the introduction)
the maximal Cauchy development
of $T^2$ symmetric $C^\infty$ initial data on $T^3$ for the
Einstein-Vlasov equations\footnote{The area of the group orbits
need not be constant on the initial data surface.  The hypotheses
concerning the Vlasov matter are stated later in this section.}
is isometric to $(t_0,\infty) \times T^3$
with $C^\infty$ metric,
\begin{eqnarray}
\label{metric}
g &=& e^{2(\nu-U)}(-\alpha \, dt^2+
d\theta^2)+e^{2U}[dx+A \, dy+(G+AH) \, d\theta]^2\nonumber\\
& & +e^{-2U}t^2[dy+H \, d\theta]^2,
\end{eqnarray}
for some nonnegative number $t_0$.
All the metric functions are periodic in $\theta$ with period
one and independent of $x$ and $y$ so that $X = \partial_x$ and
$Y = \partial_y$ are commuting Killing vector fields.  The area
of the symmetry orbits is $t$.  (The result in \cite{a1} is
similar, but with an additional assumption made, that
$G$ and $H$ vanish.)

The goal of the present work is to show that $t_0 = 0$, that
the area of the group orbits goes to zero at the end of the
maximal Cauchy development, in the contracting direction.
The strategy for showing that $t_0 = 0$ will be to
show that, given a $C^\infty$ solution to the Einstein-Vlasov
equations with metric~(\ref{metric})
on $(t_0,\infty) \times T^3$, and a nonvanishing number of
Vlasov particles (see below), then
for any $t_p \in [t_0,\infty)$ such that $t_p > 0$,
the spacetime extends to $t_p$.  It will
be convenient to fix $t_p \in [t_0,\infty)$
such that $t_p > 0$, and to fix some arbitrary
$t_i \in (t_p,\infty)$.

There are two twist quantities
associated with the commuting Killing vector fields,
$J = \epsilon_{\mu \rho \sigma \tau}
X^\mu Y^\rho \nabla^\sigma X^\tau$ and
$K =
\epsilon_{\mu \rho \sigma \tau} X^\mu Y^\rho \nabla^\sigma Y^\tau$.
If the orientation is such that
$\{\partial_t,\partial_\theta,\partial_x,\partial_y\}$ is
right handed, then these twist quantities are related to the
metric functions as follows,\footnote{With this choice of
orientation, $K$ as defined here is related to
$K$ as defined in \cite{BCIM} and \cite{IW} by change of
sign.  $J=0$ in \cite{BCIM} and \cite{IW}.}
\begin{equation}
\label{JKtometric}
J = - \frac{t e^{-2 \nu + 4U}}{\sqrt{\alpha}}(G_t + A H_t)
\hspace{20pt} \mbox{and} \hspace{20pt}
K = A \, J - \frac{t^3 e^{-2 \nu}}{\sqrt{\alpha}}H_t.
\end{equation}

Consider massive Vlasov particles all of the same mass.  Without
loss of generality, the mass can then be normalized to one.
Let the velocity vector of the Vlasov matter be
future directed (with the future designated as increasing $t$).
Let the quantities, $v_\mu$, denote the components of
the velocity vector in the frame
\begin{equation}
\label{frame}
\{dt,\;d \theta,\; dx + G \, d\theta,\; dy + H \, d\theta\}.
\end{equation}
With this choice of frame, $v_2$ and $v_3$ are constant along
geodesics, and from $v_\mu v^\mu = -1$,
\begin{equation}
\label{v0}
v_0 = - \sqrt{\alpha e^{2 \nu - 2 U} + \alpha v_1^2
+ \alpha e^{2 \nu - 4 U} v_2^2
+ \alpha t^{-2} e^{2 \nu} (v_3 - A v_2)^2}.
\end{equation}
The Vlasov matter distribution function, $f$, is in general
a nonnegative function of position and velocity (or momentum).
Suppose that $C^\infty$ initial data is given
for the Einstein-Vlasov
equations\footnote{See \cite{r1,a2,r2} for statements of the
Einstein-Vlasov initial value problem.}
on some $T^2$ symmetric initial three-torus, with
a nonvanishing number of Vlasov particles
(see the paragraph following lemma~\ref{continuityequation}
below), with
the matter distribution function invariant under the action of
the symmetry group and with the support of $f$ bounded initially.
From \cite{ARW} it is known
that $f$ is $C^\infty$ on $(t_0,\infty) \times T^3 \times R^3$,
and the support of $f$ is bounded at each $t \in (t_0,\infty)$.
The Vlasov equation is that $f$ should be constant along
timelike geodesics.
In coordinates $(t,\theta,x,y,v_1,v_2,v_3)$,
the Vlasov equation takes the form
\newpage
\begin{eqnarray}
\label{ft}
\frac{\partial f}{\partial t}  & = &
\frac{\partial v_0}{\partial v_1} \,
\frac{\partial f}{\partial \theta}
- \bigg\{ \frac{\partial v_0}{\partial \theta}
\nonumber \\ &&\; \; +
\frac{\sqrt{\alpha} e^{2 \nu}}{t^3} (K - A \, J)(v_3 - A \, v_2)
+ \frac{\sqrt{\alpha} e^{2 \nu - 4U}}{t} J v_2 \bigg\}
\frac{\partial f}{\partial v_1},
\end{eqnarray}
with $f_x = 0$,
$f_y = 0$ and $f$ periodic in $\theta$ with period one.

\begin{lemma}
\label{maxfconstant}
There exists a number $C_1>0$ such that
$f \leq C_1$ on $(t_p,t_i] \times T^3 \times R^3$.
\end{lemma}
\textit{Proof:}
This follows from $f$ being constant on timelike geodesics.
\hfill$\Box$

The second lemma follows from the continuity equation,
$\nabla_\mu N^\mu = 0$, which holds generally for Vlasov matter,
with $N^\mu$ the matter current density.  In the frame (\ref{frame}),
the matter current density is $$N^\mu = \frac{\sqrt{\alpha}}{t}
\int_{R^3} \frac{f v^\mu}{|v_0|} \, dv_1\,dv_2\,dv_3.$$

\begin{lemma}
\label{continuityequation}
The quantity
$C_2=\int_{S^1} (\int_{R^3} f \; dv_1\,dv_2\,dv_3)\,d\theta$
is constant in time.
\end{lemma}
\textit{Proof:}
This lemma can be obtained directly from the continuity equation.
Similarly, it can be obtained from equation (\ref{ft}), as follows.
\begin{eqnarray}
\frac{d C_2}{dt}& = &
\label{before}
\int_{S^1} \bigg(\int_{R^3}
\bigg\{ \frac{\partial v_0}{\partial v_1} \,
\frac{\partial f}{\partial \theta}
- \Big[ \frac{\partial v_0}{\partial \theta}
+ \frac{\sqrt{\alpha} e^{2 \nu}}{t^3} (K - A \, J)(v_3 - A \, v_2)
\nonumber \\ && \; \;
+ \frac{\sqrt{\alpha} e^{2 \nu - 4U}}{t} J v_2 \Big] 
\frac{\partial f}{\partial v_1} \bigg\}
\, dv_1\,dv_2\,dv_3\bigg)\,d\theta, \\ &=&
\label{after}
\int_{S^1} \bigg(\int_{R^3}\bigg\{ \frac{\partial v_0}{\partial v_1}
\, \frac{\partial f}{\partial \theta}
+ \frac{\partial^2 v_0}{\partial v_1 \, \partial \theta}
f \bigg\} \, dv_1\,dv_2\,dv_3\bigg)\,d\theta, \\ &=&
\int_{S^1} \bigg(\int_{R^3} \frac{\partial}{\partial \theta}
\left(\frac{\partial v_0}{\partial v_1} f \right)
\, dv_1\,dv_2\,dv_3\bigg)\,d\theta, \nonumber \\ &=&0.
\end{eqnarray}
Integration by parts leads from line (\ref{before}) to
line (\ref{after}).
\hfill$\Box$

The property that the number of Vlasov particles is nonvanishing
is equivalent to $C_2 \neq 0$, which is assumed throughout
the rest of this work.

Since $f$ doesn't grow arbitrarily large, since the integral of
$f$ is a constant in time, and since $f$, $v_2$ and $v_3$ are
all constant along timelike geodesics,
$\int_{S^1} ( \int_{R^3}
f |v_1|\; dv_1 \, dv_2 \, dv_3 ) \, d\theta $
does not become arbitrarily small on $(t_p,t_i]$, as shown by the
following lemma.

\begin{lemma}
\label{intf|v_1|}
There exists $C_3 > 0$ such that
$\int_{S^1} (\int_{R^3}
f |v_1|\; dv_1 \, dv_2 \, dv_3 )\, d\theta > C_3 $
on $(t_p,t_i]$.
\end{lemma}
\textit{Proof:}
Fix positive numbers $\bar v_2$ and $\bar v_3$ such that
the support of $f$ at $t_i$ is contained in
$S^1 \times R^1 \times [-\bar v_2,\bar v_2]
\times [-\bar v_3,\bar v_3]$.  At any $t \in (t_p,t_i]$,
the support of $f$ is contained within this same set.
This is because $f$, $v_2$ and $v_3$
are constant along timelike geodesics, so that if $f$ vanishes for
all $\theta$ and $v_1$ at some $(t_i,v_2,v_3)$ then $f$ vanishes
for all $\theta$ and $v_1$ at $(t,v_2,v_3)$ for all $t \in (t_p,t_i]$.
Fix $\delta > 0$ small enough so that
$b = C_2 - 8 \delta \bar v_2 \bar v_3 C_1$ is positive.
From the preceding paragraph and lemma \ref{maxfconstant},
\begin{equation}
\label{insidesmall}
\int_{S^1} \bigg(
\int_{R^2} \left(\int_{-\delta}^\delta f \; dv_1 
\right) dv_2 \, dv_3 
\bigg)\, d\theta \leq 8 \delta \bar v_2 \bar v_3 C_1.
\end{equation}
Now using lemma \ref{continuityequation},
$$\int_{S^1} 
\bigg( \int_{R^2} \left(\int_{|v_1| > \delta} f \; dv_1 
\right) dv_2 \, dv_3 \bigg) \, d\theta \geq b.$$
To obtain lemma \ref{intf|v_1|} calculate
\begin{eqnarray}
\int_{S^1}
\bigg( \int_{R^3} f |v_1|\; dv_1 \, dv_2 \, dv_3
\bigg) \, d\theta
& = & 
\int_{S^1}
\bigg( \int_{R^2} \left(\int_{-\delta}^\delta f |v_1|\; dv_1 
\right) dv_2 \, dv_3 \bigg) \, d\theta \nonumber \\ &&\;\;
+ \int_{S^1}
\bigg( \int_{R^2} \left(\int_{|v_1| > \delta} f |v_1|\; dv_1 
\right) dv_2 \, dv_3
\bigg) \, d\theta \nonumber \\ &\geq&
\delta \int_{S^1}
\bigg( \int_{R^2} \left(\int_{|v_1| > \delta} f \; dv_1 
\right) dv_2 \, dv_3
\bigg) \, d\theta \nonumber \\ &\geq&
\delta b,
\end{eqnarray}
and let $C_3 = \delta b$.
\hfill$\Box$

In the course of the proof of the main result, a coordinate
transformation,
$(t,\theta,v_1,v_2,v_3) \rightarrow
(s,\phi, w_1,w_2,w_3)$,
will be used, with $t = s$ (letting $s_i = t_i$ and
$s_p = t_p$), $v_2 = w_2$, $v_3 = w_3$,
$\partial \theta / \partial s = - \partial v_0 / \partial v_1$,
$$\frac{\partial v_1}{\partial s}
= \frac{\partial v_0}{\partial \theta} + \sqrt{\alpha} e^{2\nu}
\left\{\frac{(K - A \, J)(v_3 - A\, v_2)}{t^3} +
\frac{e^{-4U} J \, v_2}{t}\right\},$$
$\theta(s_i,\phi, w_1,w_2,w_3) = \phi$
and $v_1(s_i,\phi, w_1,w_2,w_3) = w_1$.  The new
coordinates, $(\phi,w_1,w_2,w_3)$, are constant on timelike
geodesics and therefore $f_s = 0$, which is why the coordinate 
transformation is useful.

\begin{lemma}
\label{det}
The Jacobian determinant of the transformation
$(t,\theta,v_1,v_2,v_3) \rightarrow
(s,\phi, w_1,w_2,w_3)$,
equals one on $(s_p,s_i] \times S^1 \times R^3$.
\end{lemma}
\textit{Proof:}
The Jacobian determinant of the coordinate transformation is
$$\eta = \frac{\partial \theta}{\partial \phi}
\frac{\partial v_1}{\partial w_1}
- \frac{\partial \theta}{\partial w_1}
\frac{\partial v_1}{\partial \phi}.$$
On $\{s_i\} \times S^1 \times R^3$, $\eta = 1$, and 
on $(s_p,s_i] \times S^1 \times R^3$,
\begin{eqnarray}
\label{detconstant}
\eta_s & = & \frac{\partial}{\partial \theta} 
\bigg(\frac{\partial \theta}{\partial s}\bigg)
\frac{\partial \theta}{\partial \phi}
\frac{\partial v_1}{\partial w_1}
+ \frac{\partial}{\partial v_1} 
\bigg(\frac{\partial \theta}{\partial s}\bigg)
\frac{\partial v_1}{\partial \phi}
\frac{\partial v_1}{\partial w_1}
\nonumber \\ &&\;\:
+ \frac{\partial \theta}{\partial \phi}
\frac{\partial}{\partial \theta} 
\bigg(\frac{\partial v_1}{\partial s}\bigg)
\frac{\partial \theta}{\partial w_1}
+ \frac{\partial \theta}{\partial \phi}
\frac{\partial}{\partial v_1} 
\bigg(\frac{\partial v_1}{\partial s}\bigg)
\frac{\partial v_1}{\partial w_1}
\nonumber \\ &&\;\:
- \frac{\partial}{\partial \theta} 
\bigg(\frac{\partial \theta}{\partial s}\bigg)
\frac{\partial \theta}{\partial w_1}
\frac{\partial v_1}{\partial \phi}
- \frac{\partial}{\partial v_1} 
\bigg(\frac{\partial \theta}{\partial s}\bigg)
\frac{\partial v_1}{\partial w_1}
\frac{\partial v_1}{\partial \phi}
\nonumber \\ &&\;\:
- \frac{\partial \theta}{\partial w_1}
\frac{\partial}{\partial \theta} 
\bigg(\frac{\partial v_1}{\partial s}\bigg)
\frac{\partial \theta}{\partial \phi}
- \frac{\partial \theta}{\partial w_1}
\frac{\partial}{\partial v_1} 
\bigg(\frac{\partial v_1}{\partial s}\bigg)
\frac{\partial v_1}{\partial \phi}.
\end{eqnarray}
From
$$\frac{\partial}{\partial \theta}
\bigg(\frac{\partial \theta}{\partial s}\bigg)
+ \frac{\partial}{\partial v_1}
\bigg(\frac{\partial v_1}{\partial s}\bigg) = 0,$$
and cancellation of terms in (\ref{detconstant}),
it follows that $\eta_s = 0$.
\hfill$\Box$

In general, coordinates that are constant on timelike
geodesics are valid on the maximal Cauchy development
of an initial data surface.  Here one can see this also
directly from lemma~\ref{det} combined with the smoothness
of the metric functions appearing in (\ref{metric}) on the
maximal Cauchy devolopment.

\section{Einstein's equations}
\label{ee}

With choice of frame (\ref{frame}), the stress-energy tensor
due to the Vlasov matter is
\begin{equation}
\label{stressenergy}
T_{\mu \rho} = \frac{\sqrt{\alpha}}{t}
\int_{R^3} \frac{f v_\mu v_\rho}{|v_0|}
dv_1 \, dv_2 \, dv_3.
\end{equation}
\newpage
\noindent Einstein's equations, $G_{\mu \rho} = T_{\mu \rho}$,
for the metric (\ref{metric})
with source (\ref{stressenergy}) are
satisfied if
\begin{eqnarray}
\label{nut}
\nu_t &=& t\,\left\{U_t^2 + \alpha U_\theta^2
+\frac{e^{4U}}{4t^2}(A_t^2 + \alpha A_\theta^2)\right\}
+ \frac{\alpha e^{2 \nu - 4 U}}{4t}J^2\nonumber\\&&
+ \frac{\alpha e^{2 \nu}}{4t^3}(K - A \, J)^2
+ \sqrt{\alpha}\int_{R^3} f |v_0|\; dv_1 \, dv_2 \, dv_3, \\
\label{nutheta}
\nu_\theta &=& 2t\left(U_t U_\theta + \frac{e^{4U}}{4t^2}
A_t A_\theta \right)
- \frac{\alpha_\theta}{2 \alpha}
- \sqrt{\alpha} \int_{R^3} f \, v_1 \; dv_1 \, dv_2 \, dv_3, \\
\label{alphat}
\frac{\alpha_t}{\alpha} &=& 
-\frac{\alpha e^{2\nu - 4U} J^2}{t}
-\frac{\alpha e^{2\nu} (K - A \, J)^2}{t^3}\nonumber\\&&
- 2\alpha^{3/2}e^{2\nu}\!\int_{R^3}\frac{f (e^{- 2U} \! \!
+ e^{- 4 U} v_2^2 + t^{-2} (v_3 - A v_2)^2)}
{|v_0|}dv_1 \, dv_2 \, dv_3, \\
\label{Jt}
J_t & = & -2 \alpha \int_{R^3}
\frac{f v_1 v_2 }{ |v_0| }dv_1 \, dv_2 \, dv_3, \\
\label{Jtheta}
J_\theta & = & 2 \int_{R^3} f v_2 \; dv_1 \, dv_2 \, dv_3, \\
\label{Kt}
K_t & = & -2 \alpha \int_{R^3}
\frac{f v_1 v_3 }{|v_0|}dv_1 \, dv_2 \, dv_3, \\
\label{Ktheta}
K_\theta & = & 2 \int_{R^3} f v_3 \; dv_1 \, dv_2 \, dv_3, \\
\label{Utt}
U_{tt} & = & \alpha U_{\theta\theta} - \frac{U_t}{t}
+ \frac{\alpha_t U_t}{2\alpha} + \frac{\alpha_\theta U_\theta}{2}
+ \frac{e^{4U}}{2t^2}\left({A_t}^2 - \alpha {A_\theta}^2\right)
\nonumber\\&&
+ \frac{\alpha e^{2 \nu - 4 U}}{2t^2}J^2
+ \frac{\alpha^{3/2}e^{2\nu-2U}}{2 t}\int_{R^3}\frac{f (1
+ 2 e^{-2  U} v_2^2)}{|v_0|}dv_1 \, dv_2 \, dv_3,\\
\label{Att}
A_{tt} & = & \alpha A_{\theta\theta} + \frac{A_t}{t}
+ \frac{\alpha_t A_t}{2\alpha} + \frac{\alpha_\theta A_\theta}{2}
- 4(A_t U_t - \alpha A_\theta U_\theta)
\nonumber\\&&
+ \frac{\alpha e^{2 \nu - 4 U}}{t^2} J (K - A \, J)
\nonumber\\&&
+ \frac{2 \alpha^{3/2} e^{2\nu-4U}}{t}\int_{R^3}
\frac{f v_2 (v_3 - A v_2)}{|v_0|}dv_1 \, dv_2 \, dv_3.
\end{eqnarray}
Since $J$ and $K$ are periodic in $\theta$,
equations~(\ref{Jtheta}) and (\ref{Ktheta}) impose
requirements on $f$.  Namely,
$\int_{S^1} \int_{R^3} f v_2 \; dv_1 \, dv_2 \, dv_3 \, d\theta = 0$
and
$\int_{S^1} \int_{R^3} f v_3 \; dv_1 \, dv_2 \, dv_3 \, d\theta = 0$.
Similarly, equation~(\ref{nutheta}) imposes the condition
\begin{equation}
\int_{S^1}\left\{ 2t\left(U_t U_\theta + \frac{e^{4U}}{4t^2}
A_t A_\theta \right)
- \sqrt{\alpha} \int_{R^3} f \, v_1 \; dv_1 \, dv_2 \, dv_3
\right\} d\theta = 0.
\end{equation}
All three of these conditions are preserved by the evolution.  That
this is so is equivalent to the integrability conditions for
equations (\ref{Jt}) and (\ref{Jtheta}), equations (\ref{Kt})
and (\ref{Ktheta}),
and equations (\ref{nut}) and (\ref{nutheta}), respectively.

It is useful to also have at hand an equation which
can be derived from (\ref{ft}) and (\ref{nut}) - (\ref{Att})
or, alternatively, can be obtained directly from Einstein's
equations, $G_{\mu \rho} = T_{\mu \rho}$,
\begin{eqnarray}
\label{nutt}
\nu_{tt} &=& \alpha \nu_{\theta\theta} +
\frac{\alpha_t \nu_t}{2\alpha} + \frac{\alpha_\theta \nu_\theta}{2}
+\frac{\alpha_{\theta \theta}}{2} - \frac{\alpha_\theta^2}{4 \alpha}
- U_t^2 + \alpha U_\theta^2
\nonumber\\&&
+ \frac{e^{4U}}{4t^2}\left({A_t}^2 - \alpha {A_\theta}^2\right)
- \frac{\alpha e^{2 \nu - 4 U} J^2}{4t^2}
- \frac{3 \alpha e^{2 \nu}(K - A \, J)^2}{4t^4}
\nonumber\\&&
- \frac{\alpha^{3/2} e^{2 \nu}}{t^3}\int_{R^3}\frac{f (v_3 - A \, v_2)^2}
{|v_0|}dv_1 \, dv_2 \, dv_3.
\end{eqnarray}

\section{Extending the spacetime}
\label{extend}

Next a number of lemmas are presented which give sufficient
control of the metric functions in the contracting direction
so that, later in this section, a slight modification of
the combination light cone/timelike geodesic arguments
used in \cite{a1,ARW} for the expanding direction
can be applied to the contracting direction.  These arguments
give control of $U$ and $A$ and their first derivatives and also
the support of $f$ (proposition~\ref{firstderivatives}),
followed by control of the first derivatives of $f$
(proposition \ref{fder}).  $C^\infty$ bounds on all the metric
functions and on $f$ follow from light cone/timelike
geodesic arguments and the Einstein-Vlasov equations.

\begin{lemma}
\label{alphanuUbound1}
For any number $b$, $\sqrt{\alpha}e^{2 \nu + b U}$
is bounded on $(t_p,t_i] \times S^1$.
\end{lemma}
\textit{Proof:}
\begin{eqnarray}
\partial_t (t^{b^2/8} \sqrt{\alpha}e^{2 \nu +b U})
&=& (t^{b^2/8} \sqrt{\alpha}e^{2 \nu + b U})
\bigg\{ 2t\bigg[(U_t + \frac{b}{4 t})^2
\nonumber \\ && \; \;
+ \alpha U_\theta^2
+\frac{e^{4U}}{4t^2}(A_t^2 + \alpha A_\theta^2)\bigg]
\nonumber \\ && \; \;
+ \sqrt{\alpha} \int_{R^3} f \left( |v_0|
+ \frac{\alpha v_1^2}{|v_0|} \right)
dv_1 \, dv_2 \, dv_3\bigg\}, \\
& \geq & 0.
\end{eqnarray}
So on $(t_p,t_i] \times S^1$,
$$\sqrt{\alpha(t,\theta)}e^{2 \nu(t,\theta) + b U(t,\theta)}
\leq \frac{t_i^{b^2/8}}{t_p^{b^2/8}}
\sqrt{\alpha(t_i,\theta)}e^{2 \nu(t_i,\theta)
+ b U(t_i,\theta)}.$$
\hfill$\Box$

\begin{lemma}
\label{alphanuAbound1}
For any positive number $r$ and any number $\lambda$,
$$\alpha^{r/2} e^{2r \nu + \lambda U} A^2$$
is bounded on $(t_p,t_i] \times S^1$.
\end{lemma}
\textit{Proof:}
An $SL(2,R)$ transformation of the Killing vectors
relates the quantity under consideration here to
quantities of the form considered in lemma \ref{alphanuUbound1}.

Consider tilded coordinates, defined by
$\partial_{\tilde t} = \partial_t$,
$\partial_{\tilde \theta} = \partial_\theta$,
$$\tilde X = \partial_{\tilde x} = a X + b Y,$$
and
$$\tilde Y = \partial_{\tilde y} = c X + d Y,$$
with $a$, $b$, $c$ and $d$ constants such that
$ad - bc = 1$.  Then the form of the metric is unchanged
from (\ref{metric}), from which it follows that
the tilded metric functions
satisfy ``tilded'' equations (\ref{nut}) - (\ref{nutt}).
The relation between the tilded metric functions
and the untilded metric functions is
\begin{eqnarray}
\label{tildeU}
e^{2 \tilde U} & = & e^{2U} (a + Ab)^2 + t^2 e^{-2U} b^2 \\
\label{tildeA}
e^{2 \tilde U} \tilde A & = & e^{2U} (a + A b) (c + A d) 
+ t^2 e^{-2U} b d \\
\label{tildealpha}
\tilde \alpha & = & \alpha \\
\label{tildenu}
\tilde \alpha e^{2 \tilde \nu - 2 \tilde U} & = &
\alpha e^{2 \nu - 2 U}
\end{eqnarray}
Note that $\tilde t = t$ and $\tilde \theta = \theta$ and
now consider $a=0$ and $b=1$.  Choose a positive number
$q < r$.  From equations (\ref{tildeU}),
(\ref{tildealpha}) and (\ref{tildenu}),
\begin{equation}
\label{boundalphanuA}
{\tilde \alpha}^{q/2} e^{2q \tilde \nu +  2(1-q) \tilde U}
= \alpha^{q/2} e^{2q \nu + 2(1-q) U} A^2
+ \alpha^{q/2} t^2 e^{2q \nu - 2(1+q) U}.
\end{equation}
Since the tilded metric functions satisfy the same equations
as the untilded metric functions, the left hand side of
equation (\ref{boundalphanuA}) is bounded on
$(t_p,t_i] \times S^1$ from lemma \ref{alphanuUbound1}.
Since the second term on the right hand side of
equation (\ref{boundalphanuA}) is positive,
$\alpha^{q/2} e^{2q \nu + 2(1-q) U} A^2$
is bounded on $(t_p,t_i] \times S^1$.
Since $\alpha^{(r-q)/2} e^{2(r-q) \nu + (\lambda-2+2q) U}$
is bounded on $(t_p,t_i] \times S^1$
from lemma~\ref{alphanuUbound1}, lemma~\ref{alphanuAbound1}
follows, by taking the product of these two bounded
quantities.
\hfill$\Box$

Next the monotonic quantity from \cite{ARW} will be recalled.
In \cite{ARW} this quantity was called $E(t)$.  The notation
here follows \cite{IW}.  Let
\begin{equation}
\label{Ehakan}
\tilde {\cal E}(t) =
\int_{S^1} \frac{\nu_t}{\sqrt{\alpha}t} \, d\theta.
\end{equation}

\begin{lemma}
\label{Emonotonic}
$d\tilde {\cal E}/dt < 0.$
\end{lemma}
\textit{Proof:}
\begin{eqnarray}
\label{dEdt}
\frac{d \tilde {\cal E}}{dt} & = &
-\int_{S^1} \Bigg\{ \frac{2}{t}\left(\frac{U_t^2}{\sqrt{\alpha}}
+\frac{e^{4U}}{4t^2}\sqrt{\alpha}A_\theta^2\right) +
\frac{\sqrt{\alpha} e^{2 \nu - 4 U}}{2 t^3} J^2
+\frac{\sqrt{\alpha}e^{2\nu}}{t^5}(K-A\,J)^2
\nonumber\\&&\;\;
+\int_{R^3}\left(\frac{f |v_0|}{t^2} + \frac{\alpha e^{2\nu}f 
(v_3-A\,v_2)^2}
{t^4 |v_0|}\right) dv_1\,dv_2\,dv_3 \Bigg\} d\theta.
\end{eqnarray}
Equation (\ref{dEdt}) follows from the evolution
equations~(\ref{alphat})
and (\ref{nutt}) and from the vanishing of the
integral over the circle of the derivative with respect to
$\theta$ of any $C^1$ function on the circle.  A nonvanishing
number of Vlasov particles ($C_2 \neq 0$)
insures that $d\tilde {\cal E}/dt$ is strictly negative.
\hfill$\Box$

\begin{lemma}
\label{Ebounded}
$\tilde {\cal E}(t)$ is bounded on $(t_p,t_i]$.
There exists a number, ${\tilde {\cal E}}_p$, satisfying
$${\tilde {\cal E}}_p =
\lim_{t \downarrow t_p} \tilde {\cal E}(t).$$
\end{lemma}
\textit{Proof:}
From equation~(\ref{dEdt}), $d\tilde {\cal E}/dt \geq 
- 4 \tilde {\cal E}/t$.
So for any $t_k \in (t_p,t_i]$,
\begin{equation}
\tilde {\cal E}(t_k) \leq \tilde {\cal E}(t_i) + \int_{t_k}^{t_i}
\frac{4\tilde {\cal E}(t)}{t} \, dt.
\end{equation}
Applying Gronwall's lemma to this inequality (as suggested
on p 353 of \cite{a1}), we obtain
\begin{equation}
\tilde {\cal E}(t_k) \leq
\tilde {\cal E}(t_i) \left(\frac{t_i}{t_k}\right)^4,
\end{equation}
for any $t_k \in (t_p,t_i]$.  Thus $\tilde {\cal E}(t_k) <
\tilde {\cal E}(t_i) (t_i/t_p)^4$
on $(t_p,t_i)$.  This bound, together with the monotonicity of
$\tilde {\cal E}(t)$, guarantees that ${\tilde {\cal E}}_p$,
as defined in the statement of the lemma, exists.
\hfill$\Box$

Let
\begin{equation}
\label{beta}
\beta = \nu + \frac{\ln \alpha}{2}.
\end{equation}
The substitution of $e^\beta$ for $\sqrt{\alpha} e^\nu$ will be made
freely throughout the rest of this section.  This is a convenience,
because of the order in which bounds are obtained.
The advantage of $\beta$ over $\nu$ is that $\alpha_\theta$ does
not appear in the expression for $\beta_\theta$ obtained from
equation~(\ref{nutheta}),
\begin{equation}
\beta_\theta = 2t\left(U_t U_\theta + \frac{e^{4U}}{4t^2}
A_t A_\theta \right)
- \sqrt{\alpha} \int_{R^3} f \, v_1 \; dv_1 \, dv_2 \, dv_3.
\end{equation}

\begin{lemma}
\label{maxmin}
There exists a positive number, $C_4$, such that
$\int_{S^1} |\beta_\theta| \, d\theta$,
$\int_{S^1} |U_\theta| \, d\theta$,
$\int_{S^1} e^{2U} |A_\theta| \, d\theta$,
$\int_{S^1} |J_\theta| d\theta$
and $\int_{S^1} |K_\theta| d\theta$ are each less than $C_4$
on $(t_p,t_i]$.
Thus, the difference between the spatial maximum and minimum values
of each of the quantities, $\beta$, $U$, $J$ and $K$ on $S^1$
(at any fixed time $t$) is less than $C_4$.
\end{lemma}
\textit{Proof:}
Using that $\alpha$ and $\tilde {\cal E}$ are increasing with
decreasing $t$,
\begin{eqnarray}
\int_{S^1} |\beta_\theta| \, d\theta &\leq& t_i
\tilde {\cal E}_p \\
\int_{S^1} |U_\theta| \, d\theta &\leq&
\frac{\sqrt{\tilde {\cal E}_p}}
{(\min_{S_1} \alpha(t_i,\theta))^{1/4}}, \\
\int_{S^1} e^{2U} |A_\theta| \, d\theta
&\leq& \frac{2 t_i \sqrt{\tilde {\cal E}_p}}
{(\min_{S_1} \alpha(t_i,\theta))^{1/4}}, \\
\int_{S^1} |J_\theta| d\theta &\leq& 2 \bar v_2 C_2,\\
\int_{S^1} |K_\theta| d\theta &\leq& 2 \bar v_3 C_2,
\end{eqnarray}
on $(t_p,t_i]$.
\hfill$\Box$

\begin{lemma}
\label{minalphabounded}
$\min_{S_1} \alpha(t,\theta)$ is bounded on
$(t_p,t_i]$.
\end{lemma}
\textit{Proof:}
For any $t \in (t_p,t_i]$, (see equations (\ref{Ehakan}) and
(\ref{nut})),
\begin{equation}
\label{boundintf|v_0|}
\int_{S^1} \int_{R^3} f |v_0|\; dv_1 \, dv_2 \, dv_3 \, d\theta
\leq t \tilde {\cal E}(t).
\end{equation}
Now considering equation (\ref{v0}),
$$\int_{S^1} \sqrt{\alpha}
\int_{R^3} f |v_1|\; dv_1 \, dv_2 \, dv_3 \, d\theta
\leq t \tilde {\cal E}(t).$$
Therefore,
$$\sqrt{\min_{S_1} \alpha(t,\theta)} \int_{S^1}
\int_{R^3} f |v_1|\; dv_1 \, dv_2 \, dv_3 \, d\theta
\leq t \tilde {\cal E}(t).$$
From lemma \ref{intf|v_1|},
$$\sqrt{\min_{S_1} \alpha(t,\theta)} C_3
\leq t \tilde {\cal E}(t),$$
with $C_3 > 0$.  So
$$\sqrt{\min_{S_1} \alpha(t,\theta)}
\leq \frac{t \tilde {\cal E}(t)}{C_3}.$$
From lemmas \ref{Emonotonic} and \ref{Ebounded},
\begin{equation}
\min_{S_1} \alpha(t,\theta)
\leq \left(\frac{t_i {\tilde {\cal E}}_p}{C_3}\right)^2
\end{equation}
on $(t_p,t_i]$.
\hfill$\Box$

\begin{lemma}
\label{bartheta}
There exists $\bar \theta \in S_1$ such that
$\alpha(t,\bar \theta)$ is bounded on $(t_p,t_i]$.
\end{lemma}

\textit{Proof:}
Suppose there existed no $\bar \theta$ such that
$\alpha(t,\bar \theta)$ were bounded on $(t_p,t_i]$.
Since $\alpha(t,\theta)$ is increasing with decreasing $t$,
this would contradict the previous lemma.
\hfill$\Box$

\begin{corollary}
\label{alphatatbartheta}
On $(t_p,t_i]$,
$$\int_{t}^{t_i} \frac{\alpha_\sigma(\sigma,\bar \theta)}
{\alpha(\sigma,\bar \theta)} \, d\sigma$$
is bounded.
\end{corollary}
\textit{Proof:}
$\ln \alpha(t,\bar \theta)$ is bounded since $\alpha(t,\bar \theta)$
is bounded and increasing with decreasing $t$.
\hfill$\Box$

In the next four lemmas control of the metric functions
is obtained which is
sufficient for using the light cone/timelike geodesic
arguments directly in terms of areal coordinates in the contracting
direction.

\begin{lemma}
\label{alphanuUbound2}
For any number $b$, $e^{\beta + b U}$
is bounded on $(t_p,t_i] \times S^1$.
\end{lemma}
\textit{Proof:}
It follows from lemma \ref{alphanuUbound1} along with
$\alpha$ being bounded on $(t_p,t_i] \times \{\bar \theta\}$
that $e^{\beta + b U}$ is
bounded on $(t_p,t_i] \times \{\bar \theta\}$.
(See equation~(\ref{beta}) for the definition of $\beta$.)
That $e^{\beta + b U}$ is
bounded on $(t_p,t_i] \times S^1$ then follows from
the boundedness of the
difference between the maximum and minimum values
of $\beta$ and $U$ on $S^1$ (see lemma \ref{maxmin}).
\hfill$\Box$

\begin{lemma}
\label{alphanuAbound2}
For any positive number $r$ and any number $b$,
$$e^{r \beta + b U} A$$
is bounded on $(t_p,t_i] \times S^1$.
\end{lemma}
\textit{Proof:}
It follows from lemma \ref{alphanuAbound1} along with
$\alpha$ being bounded on $(t_p,t_i] \times \{\bar \theta\}$
that $e^{r \beta + b U} A$ is bounded on
$(t_p,t_i] \times \{\bar \theta\}$.  Integrating
$(e^{r \beta + b U} A)_\theta$ along a $t$ = constant
path from $(t,\bar \theta)$ to $(t, \theta_2)$ leads to
\begin{eqnarray}
(e^{r \beta + b U} A)(t,\theta_2)
& = & (e^{r \beta + b U} A)(t,\bar \theta)
\nonumber \\ && \; \;
+ \int_{\bar \theta}^{\theta_2}
\bigg\{e^{r \beta + b U} A
(r \beta_\theta + b U_\theta)
\nonumber \\ && \; \;
+ e^{r \beta +(b-2) U} e^{2 U} A_\theta
\bigg\} d\theta,\\
|(e^{r \beta + b U} A)(t,\theta_2)|
& \leq & |(e^{r \beta + b U} A)(t,\bar \theta)|
\nonumber \\ && \; \;
+ \bigg|\int_{\bar \theta}^{\theta_2}
\bigg\{e^{r \beta + b U} |A|
(r |\beta_\theta| + |b U_\theta|)
\nonumber \\ && \; \;
+ e^{r \beta + (b-2) U} e^{2 U} |A_\theta| \bigg\}
d\theta \bigg|\\
& \leq & C_5 + \bigg|\int_{\bar \theta}^{\theta_2}
e^{r \beta + b U} |A|
(r |\beta_\theta| + |b U_\theta|)\, d\theta \bigg|,
\end{eqnarray}
on $(t_p,t_i]$, for some number $C_5 > 0$
(see lemmas~\ref{alphanuUbound2} and \ref{maxmin}).
From Gronwall's lemma,
\begin{equation}
|(e^{r \beta + b U} A)(t,\theta_2)|
\leq C_5 \exp\left|\int_{\bar \theta}^{\theta_2}
(r |\beta_\theta| + |b U_\theta|)\, d\theta\right|,
\end{equation}
which (using lemma~\ref{maxmin}) determines a bound for 
$e^{r \beta + b U} A$ on $(t_p,t_i] \times S^1$.
\hfill$\Box$

\begin{lemma}
\label{alphatterm1}
$$\int_t^{t_i} \max_{S^1} [e^{2 \beta(\sigma,\theta)
- 4 U(\sigma,\theta)} (J(\sigma,\theta))^2] \, d\sigma$$
is bounded on $(t_p,t_i]$.
\end{lemma}
\textit{Proof:}
From lemma \ref{maxmin},
\begin{eqnarray}
\label{intj}
\int_t^{t_i} \max_{S^1} [e^{2 \beta(\sigma,\theta)
- 4 U(\sigma,\theta)} J(\sigma,\theta))^2] \, d\sigma
&& \nonumber \\
\leq 
\int_t^{t_i} e^{6 C_4} e^{2 \beta(\sigma,\bar \theta)
-4 U(\sigma,\bar \theta)}(J(\sigma,\bar \theta)^2
+ 2 C_4 |J(\sigma,\bar \theta)| + C_4^2) d \sigma. &&
\end{eqnarray}
The right hand side of inequality~(\ref{intj}) is bounded from
corollary~\ref{alphatatbartheta}, the evolution
equation~(\ref{alphat}) and lemma~\ref{alphanuUbound2}.
\hfill$\Box$

\begin{lemma}
\label{alphatterm2}
$$\int_t^{t_i} \max_{S^1} [e^{2 \beta(\sigma,\theta)}
(K(\sigma,\theta) - A(\sigma,\theta) \, J(\sigma,\theta))^2] d\sigma$$
is bounded on $(t_p,t_i]$.
\end{lemma}
\textit{Proof:}
Integrating $(e^\beta (K -  A \, J))_\theta$ along
a $t$ = constant path from $(t,\bar \theta)$ to $(t, \theta_2)$
leads to
\begin{eqnarray}
\label{maxK-AJ}
(e^\beta |K -  A \, J|)(t,\theta_2) & \leq &
(e^\beta |K -  A \, J|)(t,\bar \theta)
+ \bigg|\int_{\bar \theta}^{\theta_2}\bigg\{
e^\beta |K -  A \, J| |\beta_\theta|
\nonumber \\ &&\; \;
+ e^\beta \int_{R^3} f |v_2| dv_1 dv_2 dv_3
\nonumber \\ &&\; \;
+ e^\beta A \int_{R^3} f |v_3| dv_1 dv_2 dv_3
\nonumber \\ &&\; \;
+ e^{\beta - 2 U} |J| e^{2U} |A_\theta|\bigg\}
d \theta \bigg| \\
&\leq &
(e^\beta |K -  A \, J|)(t,\bar \theta) + C_6
\nonumber \\ && \; \;
+ C_4 \max_{S^1}(e^{\beta - 2 U} |J|)
\nonumber \\ && \; \;
+ \bigg|\int_{\bar \theta}^{\theta_2}
e^\beta |K -  A \, J| |\beta_\theta|
d \theta\bigg|,
\end{eqnarray}
for any $(t,\theta_2) \in (t_p,t_i] \times S^1$, with
$$C_6 = C_2(\bar v_2
\sup_{(t_p,t_i] \times S^1} (e^\beta)
+ \bar v_3 \sup_{(t_p,t_i] \times S^1} (e^\beta A)).$$
From Gronwall's lemma,
\begin{eqnarray}
(e^\beta |K -  A \, J|)(t,\theta_2) & \leq &
e^{\left|\int_{\bar \theta}^{\theta_2}
|\beta_\theta| \, d\theta\right|}
\bigg\{(e^\beta |K -  A \, J|)(t,\bar \theta) + C_6
\nonumber \\ && \; \;
+ C_4 \max_{S^1}(e^{\beta - 2 U} |J|) \bigg\}.
\end{eqnarray}
Using lemma \ref{maxmin},
\begin{eqnarray}
\max_{S^1} [(e^\beta |K -  A \, J|)(t,\theta)]
& \leq & e^{C_4} \bigg\{(e^\beta
|K -  A \, J|)(t,\bar \theta) + C_6
\nonumber \\ && \; \;
+ C_4 \max_{S^1}(e^{\beta - 2 U} |J|) \bigg\}.
\end{eqnarray}
Therefore,
\begin{eqnarray}
\label{intkaj}
\int_t^{t_i} \max_{S^1} [e^{2 \beta(\sigma,\theta)} (K(\sigma,\theta)
-  A(\sigma,\theta) \,  J(\sigma,\theta))^2] \, d \sigma
& \leq & 
\nonumber \\  \; \;
e^{2 C_4}
\int_t^{t_i} \bigg\{ e^{2 \beta(\sigma,\bar \theta)}
(K(\sigma,\bar \theta)
- A(\sigma, \bar \theta) \, J(\sigma,\bar \theta))^2
\nonumber \\  \; \; +
2 C_6 (e^\beta |K -  A \, J|)(\sigma,\bar \theta)
&& \nonumber \\ \; \;
+ 2 (e^\beta |K -  A \, J|)(\sigma,\bar \theta)
C_4 \max_{S^1}(e^{\beta - 2 U} |J|)
&& \nonumber \\ \; \;
+ \bigg(C_6 + C_4 \max_{S^1}(e^{\beta - 2 U} |J|)
\bigg)^2 \, d\sigma \bigg\}.&&
\end{eqnarray}
Using that
\begin{eqnarray}
&&2 (e^\beta |K -  A \, J|)(\sigma,\bar \theta)
C_4 \max_{S^1}(e^{\beta - 2 U} |J|)
\nonumber \\
&& \;\;\;\;\;\leq
(e^{2 \beta} (K -  A \, J)^2) (\sigma,\bar \theta)
+ C_4^2 \max_{S^1}(e^{2 \beta - 4 U} J^2), \nonumber
\end{eqnarray}
the right hand side of inequality (\ref{intkaj}) is bounded from
corollary~\ref{alphatatbartheta},
the evolution equation~(\ref{alphat}) and
lemma~\ref{alphatterm1}.
\hfill$\Box$

\begin{proposition}
\label{firstderivatives}
The functions $U$ and $A$ and their first derivatives are bounded
on $(t_p,t_i] \times S^1$.  The support of $f$ is bounded on
$(t_p,t_i]$.
\end{proposition}
\textit{Proof:}
The proof of this proposition (which has to do with the contracting 
direction) is a slight modification
of the argument used for the expanding
direction in \cite{a1} and \cite{ARW}.  Let
\begin{eqnarray}
\label{E}
E &=& \frac{1}{2} \{U_t^2+\alpha U_\theta^2
+\frac{e^{4U}}{4t^2} (A_t^2 + \alpha A_\theta^2 )\},\\
\label{P}
P &=& \sqrt{\alpha}
(U_t U_\theta+\frac{e^{4U}}{4t^2} A_t A_\theta).
\end{eqnarray}
Since $E \pm P$ are both the sum of squares, $|P| \leq E$.
Set
\begin{equation}
\label{nullderivatives}
(\partial_t \mp \sqrt{\alpha}\partial_\theta)
(E \pm P) = L_\pm. 
\end{equation}
\newpage
\noindent A straightforward calculation shows that
\begin{eqnarray}
\label{lpm}
L_\pm & = & - \frac{1}{t}\left({U_t^2}
+\frac{e^{4U}}{4t^2}\alpha A_\theta^2\right)
+\frac{\alpha_t}{\alpha}(E \pm P) \mp \frac{P}{t}
\nonumber \\ && \; \;
+ (U_t \pm \sqrt{\alpha} U_\theta)
\Bigg\{ \frac {e^{2 \beta - 4 U}}{2t^2} J^2
\nonumber \\ && \; \;
+ \frac{\sqrt{\alpha} e^{2 \beta - 2 U}}{2t} \int_{R^3}\frac{f (1
+ 2 e^{-2  U} v_2^2)}{|v_0|}dv_1 \, dv_2 \, dv_3 \Bigg\}
\nonumber \\&& \; \;
+ \frac{e^{2 U}}{2t}(A_t \pm \sqrt{\alpha} A_\theta)
\Bigg\{ \frac{e^{2 \beta - 2U}}{2 t^3} J (K - A \, J)
\nonumber \\ && \; \;
+ \frac{\sqrt{\alpha}e^{2 \beta - 2U}}{t^2} \int_{R^3}
\frac{f v_2 (v_3 - A v_2)}{|v_0|}dv_1 \, dv_2 \, dv_3 \Bigg\}.
\end{eqnarray}
Let $u_1 = \sqrt{\alpha} v_1$.
\begin{eqnarray}
\label{u1s}
(u_1^2)_s & = & \frac{\alpha_t}{\alpha} u_1^2
+ \frac{2 \sqrt{\alpha} u_1}{v_0}
\bigg\{e^{2 \beta - 2U} (\beta_\theta - U_\theta) 
\nonumber \\ && \; \; 
+ e^{2 \beta - 4 U} (\beta_\theta - 2 U_\theta) v_2^2
+ \frac{e^{2 \beta}}{t^2}
(v_3 - A v_2) [(v_3-Av_2) \beta_\theta
\nonumber \\ & & \; \;
 - A_\theta v_2]\bigg\}
+ \frac{2 e^{2 \beta} u_1}{t} \left\{\frac{(K-A\,J) (v_3
- A v_2)}{t^2} + e^{-4U} J v_2 \right\}
\end{eqnarray}
The advantage of $u_1$ over $v_1$ is that $\alpha_\theta$ does
not appear in equation~(\ref{u1s}).  Let
$\bar u_1(t) = \sup \{\sqrt{\alpha} |v_1| : f(\hat t,\theta,v_1,v_2,v_3)
\neq 0 \hspace{5pt} \mbox{for some} \hspace{5pt}
(\hat t,\theta,v_2,v_3) \in [t,t_i] \times S^1 \times R^2 \}$.
Analogously to
the expanding direction in \cite{a1} and \cite{ARW}
(but here considering the contracting direction),
the strategy to prove the proposition is to show that
$\max_{S^1} E(t, \theta) + \bar u_1^2(t)$ is bounded
on $(t_p,t_i]$.  Let
\begin{equation}
\Psi(t) = \left\{\begin{array}{ll}
\max_{S^1} E(t, \theta) + \bar u_1^2(t) & \mbox{if}
\max_{S^1} E(t, \theta) + \bar u_1^2(t) >1 \\
1 & \mbox{if}
\max_{S^1} E(t, \theta) + \bar u_1^2(t) \leq 1. \\
\end{array} \right.
\end{equation}
If
\begin{equation}
\label{boundpsi1}
\Psi(t) \leq C_7 + \int_t^{t_i} h(\sigma) 
\Psi(\sigma) \ln \Psi(\sigma) \, d\sigma
\end{equation}
on $(t_p,t_i]$, for some number $C_7 > 1$ and function
$h(t) \geq 0$, it follows
(since $\Psi \geq 1$) that
$$\frac {h(t) \Psi(t) \ln \Psi(t)}
{(C_7 + \int_t^{t_i} h(\sigma) \Psi(\sigma)
\ln \Psi(\sigma) \, d\sigma)
\ln(C_7 + \int_t^{t_i} h(\sigma) \Psi(\sigma)
\ln \Psi(\sigma) \, d\sigma)} \leq h(t).$$
Integrating both sides of this inequality and then
exponentiating twice and using inequality~(\ref{boundpsi1})
one finds that
\begin{equation}
\label{boundpsi3}
\Psi(t) \leq
C_7^{\exp\left(\int_t^{t_i}h(\sigma) \, d\sigma\right)}
\end{equation}
on $(t_p,t_i]$.  It will be shown that $\Psi$ satisfies
inequality (\ref{boundpsi1}) with
the quantity $\int_t^{t_i} h(\sigma) \, d\sigma$
bounded on $(t_p,t_i]$, so that inequality~(\ref{boundpsi3})
determines a bound for $\Psi$ on $(t_p,t_i]$.

First consider $\max_{S^1} E(t, \theta)$.
Consider an arbitrary point
$(t_k,\theta_k) \in (t_p,t_i] \times S^1$.
Let $\gamma_\pm$ be the integral curves of
$\partial_t \pm \sqrt{\alpha} \partial_\theta$ starting from the point
$(t_k,\theta_k)$ and extending to the surface $t = t_i$.
Let $(t_i,\theta_\pm)$ denote the endpoints of
$\gamma_\pm$ lying on the surface $t=t_i$.  Then
\begin{equation}
\int_{\gamma_-}(\partial_t - \sqrt{\alpha}\partial_\theta)
(E + P) +
\int_{\gamma_+}(\partial_t + \sqrt{\alpha}\partial_\theta)
(E - P) = \int_{\gamma_-} L_+ + \int_{\gamma_+} L_-.
\end{equation}
So
\begin{eqnarray}
E(t_k,\theta_k) &=& \frac{1}{2}\bigg\{E(t_i,\theta_+)
- P(t_i,\theta_+) + E(t_i,\theta_-) + P(t_i,\theta_-)
\nonumber \\ && \; \;
- \int_{\gamma_+} L_- - \int_{\gamma_-} L_+ \bigg\}, \nonumber \\
\label{boundEa}
&\leq& E(t_i,\theta_+) + E(t_i,\theta_-) \nonumber \\
&& \; \; + \frac{1}{2}\left\{\int_{\gamma_+} |L_-|
+ \int_{\gamma_-}|L_+|\right\}.
\end{eqnarray}

Consider $L_\pm$ (see equation~(\ref{lpm})).  The terms
in the brackets satisfy
\begin{equation}
\frac {e^{2 \beta - 4 U}}{2t^2} J^2
+ \frac{\sqrt{\alpha} e^{2 \beta - 2 U}}{2t} \int_{R^3}
\frac{f (1 + 2 e^{-2  U} v_2^2)}{|v_0|}dv_1 \, dv_2 \, dv_3
\leq |\frac{\alpha_t}{2 \alpha t}|,
\end{equation}
\begin{equation}
\frac{e^{2 \beta - 2U}}{2 t^3} J (K - A \, J)
+ \frac{\sqrt{\alpha} e^{2 \beta-2U}}{t^2} \int_{R^3}
\frac{f v_2 (v_3 - A v_2)}{|v_0|}dv_1 \, dv_2 \, dv_3
\leq |\frac{\alpha_t}{2 \alpha t}|.
\end{equation}
The sum of the factors in front of the brackets satisfies
\begin{eqnarray}
U_t \pm \sqrt{\alpha} U_\theta
+ \frac{e^{2U}}{2t}(A_t \pm \sqrt{\alpha} A_\theta)
&\leq& (U_t \pm \sqrt{\alpha} U_\theta)^2
+ \frac{e^{4U}}{4t^2}(A_t \pm \sqrt{\alpha} A_\theta)^2
+ \frac{1}{2}
\nonumber \\ 
&\leq& 2(E \pm P) + \frac{1}{2}
\nonumber \\ 
&\leq& 4 E + \frac{1}{2}.
\end{eqnarray}
Thus
\begin{equation}
\label{boundlpm}
|L_\pm| \leq |\frac{\alpha_t}{\alpha}|(2E + \frac{2E}{t}
+ \frac{1}{4t}) + \frac{3E}{t}
\end{equation}

The first two terms in $\alpha_t/\alpha$ (see equation~(\ref{alphat}))
are treated in lemmas~\ref{alphatterm1} and \ref{alphatterm2}.
Changing variables from $v_1$ to $u_1$, the third term satisfies
\begin{eqnarray}
&& 2 e^{2\beta- 2U}
\int_{R^3}\frac{f (1+ e^{- 2 U} v_2^2 + t^{-2} e^{2U}(v_3 - A v_2)^2)}
{|v_0|}du_1 \, dv_2 \, dv_3
\nonumber \\ &\leq&
\label{comment}
2 e^{2\beta- 2U}
\int_{R^3}\frac{f (1+ e^{- 2 U} v_2^2 + t^{-2} e^{2U}(v_3 - A v_2)^2)}
{\sqrt{e^{2\beta- 2U} + u_1^2}}du_1 \, dv_2 \, dv_3 \\ &\leq&
8 C_1 (e^{\beta- U} + e^{\beta - 3 U} \bar v_2^2
+ \frac{e^{\beta + U}}{t^2} (\bar v_3 + |A| \bar v_2)^2)
\bar v_2 \bar v_3
\int_{-\bar u_1}^{\bar u_1}\frac{du_1}
{\sqrt{1 + e^{-2\beta + 2 U} u_1^2}}
\nonumber \\ &\leq&
\label{alphatterm3}
16 C_1 \Big(e^{\beta- U}
+ e^{\beta - 3 U} \bar v_2^2 \nonumber \\ && \; \;
+ \frac{e^{\beta + U}}{t^2} (\bar v_3 + |A| \bar v_2)^2\Big)
\bar v_2 \bar v_3 \Big(e^{- 1}+e^{\beta- U}
\ln\big(\bar u_1 + \sqrt{e^{2\beta - 2 U} +  \bar u_1^2}\big)
\Big).
\end{eqnarray}
One of the factors, $e^{\beta -U}$, appearing in front of
the integral in line~(\ref{comment}) is used to control
the terms appearing in the numerator and the other
has been absorbed into the integral in the following line,
to control it, since it has not yet been shown that 
$e^{\beta -U}$ is bounded away from zero.
Now using lemmas~\ref{alphanuUbound2} and
\ref{alphanuAbound2}, and inequalities~(\ref{boundEa}),
(\ref{boundlpm}) and (\ref{alphatterm3}),
\begin{eqnarray}
\label{boundEb}
\max_{S^1} E(t_k,\theta) & \leq & 2  \max_{S^1} E(t_i,\theta)
\nonumber \\ && \; \; 
+ \int_{t_k}^{t_i}\Bigg\{ \bigg\{
\frac{\max_{S^1} [e^{2 \beta(t,\theta)
- 4 U(t,\theta)} (J(t,\theta))^2]}{t}
\nonumber \\ && \; \; 
+ \frac{ \max_{S^1} [e^{2 \beta(t,\theta}
(K(t,\theta) - A(t,\theta) \, J(t,\theta))^2]} {t^3}
\nonumber \\ && \; \; 
+ C_8 (e^{-1} + e^{\beta - U}
\ln (\max\{\bar u_1,e^{\beta - U}\})
\bigg\} (2E + \frac{2E}{t_p} + \frac{1}{4 t_p})
\nonumber \\ && \; \; 
+\frac{3E}{t_p} \Bigg\} dt,
\end{eqnarray}
for some positive number $C_8$.

Next consider $\bar u_1^2$.  Integrating along
geodesics using equation~(\ref{u1s}), treating
$\alpha_t/\alpha$ as above, and using that
$|u_1| < |v_0|$, that
\begin{equation}
\sqrt{\alpha}|\beta_\theta| \leq
2 t E + 4 \bar u_1^2 \bar v_2 \bar v_3 C_1,
\end{equation}
that
$\sqrt{\alpha} |U_\theta| \leq 2E + 1$ and
that $\sqrt{\alpha} e^{2U} |A_\theta|/t \leq 2 E + 1$,
\begin{eqnarray}
\label{boundu1}
(u_1(t_k,\theta_k))^2 & \leq &
(\bar u_1(t_i))^2
+ \int_{t_k}^{t_i} \Bigg\{
\bigg\{\frac{\max_{S^1} [e^{2 \beta(t,\theta)
- 4 U(t,\theta)} (J(t,\theta))^2]}{t}
\nonumber \\ && \;\;
+\frac {\max_{S^1} [e^{2 \beta(t,\theta)}
(K(t,\theta) - A(t,\theta) \, J(t,\theta))^2]}{t^3}
\nonumber \\ && \;\;
+ C_8 (e^{-1} + e^{\beta - U}
\ln (\max\{\bar u_1,e^{\beta - U}\})
\bigg\} (\bar u_1(t))^2 \nonumber \\
&& \; \; + C_9(E + \bar u_1^2 + 1)
\bigg\{\max_{S^1}e^{2 \beta - 2U} + 
(\max_{S^1}e^{2 \beta - 4 U}) \bar v_2^2 
\nonumber \\ & & \; \;
+ \frac{[(\max_{S^1}e^{\beta}) \bar v_3
+ (\max_{S^1}e^{\beta}|A|)\bar v_2]^2}{t_p^2}
\nonumber \\ & & \; \;
+ \frac{[(\max_{S^1}e^{2\beta - 2U}) \bar v_3
+ (\max_{S^1}e^{2\beta-2U}|A|)\bar v_2]
\bar v_2}{t_p}
\nonumber \\ && \; \;
+ \frac{\max_{S^1}(e^{\beta}|K-A\,J|)
[(\max_{S^1}(e^{\beta}) \bar v_3
+ \max_{S^1}(e^{\beta}|A|) \bar v_2]}{t_p^3}
\nonumber \\ && \; \;
+ \frac{(\max_{S^1} e^{\beta-2U}) \max_{S^1}(e^{\beta-2U} |J|)
\bar v_2}{t} \bigg\} \Bigg\} dt
\end{eqnarray}
for some positive number $C_9$.
From (\ref{boundEb}) and (\ref{boundu1}), and
using lemmas~\ref{alphatterm1} and \ref{alphatterm2}, it follows that
$\Psi(t)$ satisfies inequality (\ref{boundpsi1}) with
$\int_t^{t_i} h(\sigma) \, d\sigma$
bounded on $(t_p,t_i]$, so that inequality~(\ref{boundpsi3})
determines a bound for $\Psi$ on $(t_p,t_i]$.
That the support of $f$ and
$U_t$ are bounded on $(t_p,t_i] \times S^1$ follows immediately.
Therefore $U$ is bounded and from this follows the bound on $A_t$
and therefore $A$.  The bounds on the derivatives with respect to
$\theta$ follow from the bound on $E(t,\theta)$ and
the fact that $\alpha$ is increasing with decreasing $t$.
\hfill$\Box$

\begin{corollary}
$\ln \alpha$ is bounded on $(t_p,t_i] \times S^1$.
\end{corollary}

\begin{corollary}
$J$, $K$ and their first derivatives are bounded on $(t_p,t_i] \times S^1$.
\end{corollary}

\begin{corollary}
The functions $\alpha_t$, $\beta$, $\beta_t$, $\beta_\theta$,
$\nu$ and $\nu_t$ are bounded on $(t_p,t_i] \times S^1$.
\end{corollary}

\begin{proposition}
\label{fder}
The first derivatives of $f$ are bounded
on $(t_p,t_i] \times S^1 \times R^3$ and
$\alpha_\theta$ is bounded on $(t_p,t_i] \times S^1$.
\end{proposition}
\textit{Proof:}
The appropriate quantities have now been bounded so as to allow this
argument to proceed as for the expanding
direction in \cite{a1} and \cite{ARW}.  The
orthonormal frame used in \cite{a1} and \cite{ARW} is such that
the components of the velocity vector are
\begin{eqnarray}
\hat v^0 &=& -e^{-\beta+U}v_0,\\
\hat v^1 &=& \sqrt{\alpha} e^{-\beta+U}v_1,\\
\hat v^2 &=& e^{-U}v_2,\\
\hat v^3 &=& \frac{e^{U}}{t}(v_3 - A v_2).
\end{eqnarray}
Note that $\hat v^0 =
\sqrt{1 + (\hat v^1)^2 + (\hat v^2)^2 + (\hat v^3)^2}$
and that the support of $f$ is bounded in terms of these
variables also.
Let $Q=(s,\phi,w_1,w_2,w_3)$, $R=(t,\theta,v_1,v_2,v_3)$
and $S=(t,\theta,\hat v^1,\hat v^2,\hat v^3)$.
Let $Q^0 = s$, $Q^1 = \phi$, $Q^{j+i} = w_i$ and analogously for
coordinates $R$ and $S$.  Let the indices $J$, $K$ and $L$ take on
values from 0 to 4.  The goal is to determine bounds for
$\partial f/\partial R^J$.

The first step will be to determine
bounds for
$$\frac{\partial f}{\partial S^J} = \frac{\partial f}{\partial Q^K}
\frac{\partial Q^K}{\partial S^J}.$$
Since $\partial f/\partial Q^K$ is constant along timelike
geodesics, for the first step it is enough to determine bounds
for $\partial Q^K/\partial S^L$.  Since the determinants of
$\partial S/\partial R$ and $\partial R/\partial Q$
(see lemma~\ref{det}) are bounded away from zero,
the determinant of $\partial S/\partial Q$ is bounded away
from zero, so it is enough to determine
bounds for $\partial S^L/\partial Q^J$.
Let the indices $a$, $b$, $c$ and $d$ take on values from 1 to 4.
Since
$\partial S^L/\partial Q^0$ (see equations~(\ref{Ss0}) - (\ref{Ss4})
below) and $\partial S^0/\partial Q^J$ are bounded,
it is enough to determine bounds for $\partial S^a/\partial Q^b$.  Let
\begin{eqnarray}
F_a^1 & = & \frac{1}{\sqrt{\alpha}} \frac {\partial \theta}
{\partial Q^a}, \\
F_a^2 & = & \frac{\partial \hat v^1}{\partial Q^a} +
\Bigg\{ \frac{\nu_t - U_t}{\sqrt{\alpha}}{\hat v^0}
+\frac{U_t}{\sqrt{\alpha}}\frac{(\hat v^3)^2
- (\hat v^2)^2}{\hat v^0}
\nonumber \\ && \; \;
+\left(\frac{U_t}{\sqrt{\alpha}}\frac{\hat v^1}{\hat v^0} - U_\theta
\right)
\hat v^1 \frac{(\hat v^3)^2 - (\hat v^2)^2}{(\hat v^0)^2 - (\hat v^1)^2}
-\frac{e^{2U}}{t}\frac{A_t}{\sqrt{\alpha}}
\frac{\hat v^2 \hat v^3} {\hat v^0}
\nonumber \\ && \; \;
- \frac{e^{2U}}{t}
\left(\frac{A_t}{\sqrt{\alpha}}\frac{\hat v^1}{\hat v^0}
- A_\theta \right)\frac{\hat v^1 \hat v^2 \hat v^3}
{(\hat v^0)^2
- (\hat v^1)^2} \Bigg\} \frac {\partial \theta}{\partial Q^a}, \\
F_a^3 & = & \frac{\partial \hat v^2}{\partial Q^a} +
U_\theta \hat v^2 \frac {\partial \theta}{\partial Q^a}, \\
F_a^4 & = & \frac{\partial \hat v^3}{\partial Q^a} -
(U_\theta \hat v^3 - \frac{e^{2U}}{t} A_\theta \hat v^2 )
\frac {\partial \theta}{\partial Q^a}.
\end{eqnarray}
($F_b^a$ is the same as $(\Psi, Z^i)$ in \cite{a1,ARW}.)
It is straightforward to see that
$F_b^a = \Pi_c^a \partial S^c/\partial Q^b$, with
$\Pi_c^a$ and its inverse bounded.
Calculation of $\partial F_b^a/\partial s$, using
\begin{eqnarray}
\label{Ss0}
t_s & = & 1, \\
\theta_s & = & \frac{\sqrt{\alpha} \hat v^1}{\hat v^0}, \\
\hat v_s^1 & = & -(\beta_\theta - U_\theta) \sqrt{\alpha} \hat v^0
- (\beta_t - U_t - \frac{\alpha_t}{2\alpha}) \hat v^1
- \sqrt{\alpha} U_\theta
\frac{(\hat v^3)^2 - (\hat v^2)^2}{\hat v^0}
\nonumber \\ && \; \;
+ \frac{\sqrt{\alpha} e^{2 U}}{t} A_\theta
\frac{\hat v^2\hat v^3}{\hat v^0}
- \frac{e^{\beta-2U}}
{t} J \hat v^2 - \frac{e^\beta}{t^3} (K-A\,J)\hat v^3, \\
\hat v_s^2 & = &- U_t \hat v^2 - \sqrt{\alpha} U_\theta
\frac{\hat v^1 \hat v^2}{\hat v^0}, \\
\label{Ss4}
\hat v_s^3 &=& (U_t - \frac{1}{t})\hat v^3 + \sqrt{\alpha} U_\theta
\frac{\hat v^1 \hat v^3}{\hat v^0}-\frac{e^{2U}}{t}(A_t \hat v^2
+ \sqrt{\alpha} A_\theta \frac{\hat v^1 \hat v^2}{\hat v^0}),
\end{eqnarray}
shows that
$\partial F_b^a/\partial s = \Lambda_c^a \partial S^c/\partial Q^b$,
with $\Lambda_c^a$ bounded.  (All terms appearing during the course
of the calculation for which bounds have not yet been determined
are canceled by
other terms, with use of the evolution equations~(\ref{Utt}),
(\ref{Att}) and (\ref{nutt}).)  Therefore
$\partial F_b^a/\partial s = \Lambda_c^a (\Pi^{-1})^c_d F_b^d$
is sufficient to determine bounds for $F_b^a$, and therefore
$\partial S^a/\partial Q^b$.

The second step is to determine a bound for $\alpha_\theta$.
From the evolution equation~(\ref{alphat}),
\begin{eqnarray}
\label{alphattheta}
\alpha_t &=& 
-\frac{\alpha e^{2\beta - 4U} J^2}{t}
-\frac{\alpha e^{2\beta} (K - A \, J)^2}{t^3}\nonumber\\&&
- 2\alpha t e^{2\beta - 2 U} \int_{R^3}\frac{f (1
+ (\hat v^2)^2 + (\hat v^3)^2)}
{\hat v^0}d \hat v^1 \, d \hat v^2 \, d \hat v^3.
\end{eqnarray}
Taking the derivative with respect to $\theta$ of both sides
of this equation and then integrating
in time leads to an inequality of the form
\begin{equation}
|\alpha_\theta| \leq C_{10} +
\int_t^{t_i} |\alpha_\theta| \, \rho \, d\sigma,
\end{equation}
with $C_{10} > 0$ constant and $\rho(t)$ nonnegative and
bounded on $(t_p,t_i]$, so a bound for $\alpha_\theta$
is determined by Gronwall's lemma.

Finally,
\begin{equation}
\label{thirdstep}
\frac{\partial f}{\partial R^J} = \frac{\partial f}{\partial S^K}
\frac{\partial S^K}{\partial R^J}.
\end{equation}
determines bounds for $\partial f/\partial R^J$.
\hfill$\Box$

\begin{corollary}
\label{boundnutheta}
$\nu_\theta$, $\alpha_{tt}$
and $\alpha_{t \theta}$ are bounded on $(t_p,t_i] \times S^1$.
\end{corollary}

\begin{lemma}
\label{2ndderivs}
The second derivatives of $U$ and $A$ are bounded on
$(t_p,t_i] \times S^1$.
\end{lemma}
\textit{Proof:}
Let
\begin{eqnarray}
\label{nextE}
E^1 & =& \frac{1}{2} \left\{ (U_{tt})^2 + \alpha(U_{t \theta})^2
+ \frac{e^{4U}}{4 t^2}[(A_{t t})^2
+ \alpha (A_{t\theta})^2] \right\}, \\
\label{nextP}
P^1&=&\sqrt{\alpha} \left( U_{tt}U_{t \theta}
+\frac{e^{4U}}{4 t^2} A_{t t} A_{t \theta}\right).
\end{eqnarray}
A bound for $\max_{S^1} E^1(t,\theta)$ for $t \in (t_p,t_i]$
can be obtained using Gronwall's lemma
by again considering an arbitrary point
$(t_k,\theta_k) \in (t_p,t_i] \times S^1$, integrating
$(\partial_t \mp \sqrt{\alpha}\partial_\theta)(E^1 \pm P^1)$
along $\gamma_\mp$ (as defined in the proof of
proposition~\ref{firstderivatives}) and considering maxima 
on $S^1$ to obtain
\begin{equation}
\label{boundE1}
\max_{S^1} E^1(t_k,\theta) \leq C_{11}
+ \int_{t_k}^{t_i} \max_{S^1} E^1(t,\theta) \, \xi \, dt,
\end{equation}
for some positive number $C_{11}$ and some function $\xi(t)$ which
is bounded and nonnegative on $(t_p,t_i]$.
The bound for $\max_{S^1} E^1(t,\theta)$ on $(t_p,t_i]$,
determines bounds for $U_{tt}$, $U_{t\theta}$, $A_{tt}$ and
$A_{t\theta}$ on $(t_p,t_i] \times S^1$.  The bounds for
$U_{\theta \theta}$ and $A_{\theta \theta}$ then
follow from equations~(\ref{Utt}) and (\ref{Att}).
\hfill$\Box$

\begin{corollary}
The second derivatives of $J$, $K$ and $\beta$ are bounded on
$(t_p,t_i] \times S^1$.  $\nu_{t \theta}$ and $\nu_{tt}$
are bounded on $(t_p,t_i] \times S^1$.
\end{corollary}

\begin{proposition}
\label{Cinftybounds}
Let $k$ be any positive integer.
All $k$th order derivatives of the matter
distribution function and of the metric functions are bounded on
$(t_p,t_i] \times S^1 \times R^3$ and on
$(t_p,t_i] \times S^1$, respectively.
\end{proposition}
\textit{Proof:}
This follows from an inductive argument, as follows.
Let $n$ be any integer greater than one.
Suppose that bounds have been determined for all derivatives
of $\nu$, $\beta$, $\alpha$, $U$, $A$, $J$ and $K$ up to and
including order
$n$ except for $\partial^n \nu / \partial \theta^n$ and
$\partial^n \alpha / \partial \theta^n$.  In addition, suppose
that bounds have been determined for $\partial S / \partial Q$
up to and including order $n-1$.  Let
\begin{equation}
F_{(n) b_1 \cdots b_n}^a =
\frac{\partial^n S^a}{\partial Q^{b_1} \cdots \partial Q^{b_n}}
+ \frac{1}{\sqrt{\alpha}}
\frac{\partial^{n-1} \Gamma^a}{\partial \theta^{n-1}}
\frac{\partial \theta}{\partial Q^{b_1}}
\cdots \frac{\partial \theta}{\partial Q^{b_n}},
\end{equation}
with
\begin{eqnarray}
\Gamma^1 & = & -\sqrt{\alpha} \\
\Gamma^2 & = &
(\nu_t - U_t) \hat v^0 \nonumber \\ && \; \;
+ U_t \frac{(\hat v^3)^2 - (\hat v^2)^2}{\hat v^0}
+ (U_t \frac{\hat v^1}{\hat v^0} - \sqrt{\alpha} U_\theta) \hat v^1
\frac{(\hat v^3)^2 - (\hat v^2)^2} {(\hat v^0)^2 - (\hat v^1)^2}
\nonumber \\ && \; \;
- \frac{e^{2U}}{t}\left[ A_t \frac{\hat v^2 \hat v^3}{\hat v^0}
+\left(A_t \frac{\hat v^1}{\hat v^0} - \sqrt{\alpha} A_\theta\right)
\frac{\hat v^1 \hat v^2 \hat v^3} {(\hat v^0)^2
- (\hat v^1)^2}\right] \\
\Gamma^3 & = & \sqrt{\alpha} U_\theta \hat v^2 \\
\Gamma^4 & = & - \sqrt{\alpha} U_\theta \hat v^3
+ \frac{e^{2U}}{t} \sqrt{\alpha} A_\theta \hat v^2.
\end{eqnarray}
Similar to the proof of proposition~\ref{fder},
integrating $\partial_s F_{(n) b_1 \cdots b_n}^a$
along timelike geodesics determines bounds for
$\frac{\partial^n S^a}{\partial Q^{b_1} \cdots \partial Q^{b_n}}$,
followed by a bound for
$\partial^n \alpha / \partial \theta^n$ and finally bounds for
$\partial^n f/ \partial R^{J_1} \cdots \partial R^{J_n}$.
The evolution equations now determine bounds
for $\partial^n \nu / \partial \theta^n$, and the
$(n+1)$th order derivatives of $\alpha$, except for
$\partial^{n+1} \alpha / \partial \theta^{n+1}$.
Now bounds can be determined for the
$(n+1)$th order derivatives of $U$ and $A$
by carrying out $n$ different light cone arguments, with
\begin{eqnarray}
E_{mr}^n \pm P_{mr}^n & = & \frac{1}{2} \Bigg\{
\left(\frac{\partial^{n-1}U_{tt}}{\partial \theta^m
\partial t^r} \pm \sqrt{\alpha}
\frac{\partial^{n-1}U_{t\theta}}{\partial \theta^m
\partial t^r}\right)^2 \nonumber \\ & & \; \;
+ \frac{e^{4U}}{4 t^2}
\left(\frac{\partial^{n-1}A_{tt}}{\partial \theta^m
\partial t^r} \pm \sqrt{\alpha}
\frac{\partial^{n-1}A_{t\theta}}{\partial \theta^m
\partial t^r}\right)^2 \Bigg\},
\end{eqnarray}
for each choice of nonnegative integers $m$ and $r$ such that
$m + r = n-1$.
Finally, the evolution equations determine bounds for
all $(n+1)$th order derivatives of
$J$, $K$ and $\beta$, and $\nu$ except for
$\partial^{n+1} \nu / \partial \theta^{n+1}$.
\hfill$\Box$

\section{Conclusion}
\label{conclusion}

\begin{theorem}
\label{main}
The area of the $T^2$ group orbits takes on all positive values
on the maximal Cauchy development of $T^2$ symmetric initial data
for the Einstein-Vlasov equations on $T^3$
with a positive number of Vlasov
particles on the initial data surface, all of the same nonzero mass.
\end{theorem}
\textit{Proof:}
Suppose that $t_p = t_0$.
From the bound on the support of the matter distribution function
and the $C^\infty$ bounds obtained in section~\ref{extend}, it
follows that there is a
$C^\infty$ extension of the metric functions and the
matter distribution function to $t_p$
(satisfying the Einstein-Vlasov equations).  Therefore
$\{t_p\} \times S^1$ is contained in the Cauchy
development of any Cauchy surface in $(t_p,\infty) \times S^1$. 
This contradicts that the Cauchy development does not extend
to $\{t_0\} \times S^1$.  Since $t_p$ was allowed to be any
positive number in $[t_0,\infty)$, it must be the
case that $t_0 = 0$.
\hfill$\Box$

In \cite{ARW} it is shown that there is a crushing singularity as
$t \downarrow t_0$ and that there exists a foliation of the
maximal Cauchy development
by compact spatial hypersurfaces of constant mean curvature.

\section{Acknowledgments}
This work was supported by the Natural Sciences and
Engineering Research Council of Canada.  The author is grateful
to James Isenberg for comments on a draft of this text.

\section{Appendix}
\label{appendix}
An argument by Chru\'sciel concerning
vacuum $T^2$ symmetric spacetimes on $T^3$
(see pp 113-114 of \cite{Chr})
carries over to Einstein-Vlasov $T^2$ symmetric
spacetimes on $T^3$
to show that the presence of Vlasov matter guarantees that
the spacetime gradient of the area of the $T^2$ symmetry orbits is
everywhere nonvanishing.  In this appendix
the argument is recalled.  The areal form (\ref{metric})
of the metric is only valid if
the spacetime gradient of the area of
the $T^2$ symmetry orbits, $t$, is nonvanishing.
However, the conformal form of the metric,
\begin{eqnarray}
g &=& e^{2(\tilde \nu-U)}(-d\tilde t^2+d\tilde \theta^2)
+e^{2U}[d \tilde x+
A \, d \tilde y+(\tilde G+A\tilde H)
\, d \tilde \theta]^2\nonumber\\
& & +e^{-2U}t^2[d \tilde y+\tilde H \, d \tilde \theta]^2,
\end{eqnarray}
in terms of conformal coordinates
$(\tilde t, \tilde \theta, \tilde x, \tilde y)$,
is valid even if the gradient of $t$ vanishes.
(Given a spacetime with metric~(\ref{metric}), one can
set $\partial_x = \partial_{\tilde x}$ and
$\partial_y = \partial_{\tilde y}$, but to insure
zero shift it may be that
$x \neq \tilde x$ and $y \neq \tilde y$.)
The twist quantities are related to the conformal
metric functions by
\begin{equation}
\label{JKtoconfmetric}
J = - t e^{-2 \tilde \nu + 4U}
(\tilde G_{\tilde t} + A \tilde H_{\tilde t})
\hspace{20pt} \mbox{and} \hspace{20pt}
K = A \, J -
t^3 e^{-2 \tilde \nu} \tilde H_{\tilde t}.
\end{equation}

Let the frame for the velocity of the particles be
\begin{equation}
\label{confframe}
\{d \tilde t,\;d \tilde \theta,\;
d \tilde x + \tilde G \, d \tilde \theta,\;
d \tilde y + \tilde H \, d \tilde \theta\}.
\end{equation}
Then
\begin{equation}
\label{confv0}
\tilde v_0 = - \sqrt{e^{2 \tilde \nu - 2 U} + \tilde v_1^2
+ e^{2 \tilde \nu - 4 U} \tilde v_2^2
+ t^{-2} e^{2 \tilde \nu} (\tilde v_3 - A \tilde v_2)^2}.
\end{equation}

Let $h_+ = h_{\tilde t} + h_{\tilde \theta}$
and $h_- = h_{\tilde t} - h_{\tilde \theta}$
for any function, $h$.
Let
\begin{eqnarray}
B & = & t (U_+^2 + \frac{e^{4 U}}{4 t^2} A_+^2)
+\frac{e^{2 \tilde \nu-4U}}{4 t} J^2 \nonumber \\ &&
+\frac{e^{2 \tilde \nu}}{4 t^3} (K-A\,J)^2
+ \int_{R^3} f (|\tilde v_0| - \tilde v_1)
\; d \tilde v_1 \, d \tilde v_2 \, d \tilde v_3, \\
D& = & t (U_-^2 + \frac{e^{4 U}}{4 t^2} A_-^2)
+\frac{e^{2 \tilde \nu-4U}}{4 t} J^2 \nonumber \\ &&
+\frac{e^{2 \tilde \nu}}{4 t^3} (K-A\,J)^2
+ \int_{R^3} f (|\tilde v_0| + \tilde v_1) \;
d \tilde v_1 \, d \tilde v_2 \, d \tilde v_3.
\end{eqnarray}
Note that $B$ and $D$ are both strictly positive
since $|\tilde v_1| < |\tilde v_0|$ and since,
by assumption, the number of Vlasov particles
present is nonvanishing (see
lemma~\ref{continuityequation} and the paragraph
following its proof).  The following equations
can be derived from equations (5) and (6) in \cite{ARW}.
\begin{eqnarray}
(t_+)_{\tilde \theta} & = & -B + \tilde \nu_+ t_+, \\
(t_-)_{\tilde \theta} & = & D - \tilde \nu_- t_-.
\end{eqnarray}
So on any $\tilde t =$ constant surface,
\begin{eqnarray}
t_+(\tilde t,\tilde \theta) & = & t_+(\tilde t,\tilde \theta_0)
e^{\int_{\tilde \theta_0}^{\tilde \theta}
\tilde \nu_+(\tilde t,\lambda) \, d \lambda}
 - \int_{\tilde \theta_0}^{\tilde \theta}
B(\tilde t,\phi) e^{\int_\phi^{\tilde \theta}
\tilde \nu_+(\tilde t,\lambda) \, d \lambda} d \phi \\
t_-(\tilde t,\tilde \theta) & = &
t_-(\tilde t,\tilde \theta_0)
e^{-\int_{\tilde \theta_0}^{\tilde \theta}
\tilde \nu_-(\tilde t,\lambda) \, d \lambda}
+ \int_{\tilde \theta_0}^{\tilde \theta}
D(\tilde t,\phi)e^{-\int_\phi^{\tilde \theta}
\tilde \nu_-(\tilde t,\lambda) \, d \lambda} d \phi
\end{eqnarray}
Thus, if
$t_+(\tilde t,\tilde \theta_0) = 0$
for some $(\tilde t, \tilde \theta_0)$,
then $t_+(\tilde t,\tilde \theta) < 0$
for $\tilde \theta > \tilde \theta_0$.
And if $t_-(\tilde t,\tilde \theta_0) = 0$
for some $(\tilde t, \tilde \theta_0)$,
then $t_-(\tilde t,\tilde \theta) > 0$
for $\tilde \theta > \tilde \theta_0$.
This is not compatible with the periodicity of
$t_\pm$ in $\tilde \theta$,
so $t_\pm(\tilde t,\tilde \theta_0) = 0$
is not possible.  (In vacuum and if $J=0$ and
$K=0$, periodicity implies that if
$t_\pm(\tilde t,\tilde \theta_0) = 0$
for some $(\tilde t, \tilde \theta_0)$, then $t_\pm = 0$
on the whole $\tilde t=$ constant surface, which in turn
implies that $t_\pm = 0$ in the maximal Cauchy
development of the $\tilde t=$ constant surface, and that
the spacetime is flat.)

\setlength{\textheight}{600pt}

\end{document}